# Polarization-induced buildup and switching mechanisms for soliton molecules composed of noise-like-pulse transition states

Zhi-Zeng Si[1], Zhen-Tao Ju[1], Long-Fei Ren[1], Xue-Peng Wang[1], Boris A. Malomed[2,3] and Chao-Qing Dai [1*]

[1][1]*College of Optical, Mechanical and Electrical Engineering, Zhejiang A&F University, Lin'an 311300, China*

[2]*Department of Physical Electronics, Faculty of Engineering, and Center for Light-Matter Interaction, Tel Aviv University, Tel Aviv 69978, Israel*

[3]*Instituto de Alta Investigación, Universidad de Tarapacá, Casilla 7D, Arica 1000000, Chile*

**Abstract :** Buildup and switching mechanisms of solitons in complex nonlinear systems are fundamentally important dynamical regimes. Using a novel strongly nonlinear optical system, including saturable absorber MOF-253@Au and a polarization controller (PC), the work reveals a new buildup scenario for "soliton molecules (SMs)", which includes a long-duration stage dominated by the emergence of transient noise-like pulses (NLPs) modes to withstand strong disturbances arising from "turbulence" and extreme nonlinearity in the optical cavity. The switching between SMs and NLPs is controlled by the cavity polarization state. The switching involves the spectral collapse, following spectral oscillations with a variable period, and self-organization of NLPs, following energy overshoot. This switching mechanism applies to various patterns with single, paired, and clustered pulses. In multi-pulses stage, XPM (cross-phase-modulation)-induced interactions between solitons facilitate a specific mode of energy exchange between them, proportional to interaction duration, ensuring pulse stability during and after state transitions. Systematic simulations reveal effects of the PC's rotation angle and intra-cavity nonlinearity on the periodic phase transitions between the different soliton states, and accurately reproduce the experimentally observed buildup and switching mechanisms. These findings could enhance our fundamental study and points to potential uses in designing information encoding systems.

Key words: soliton dynamics, buildup and switching mechanisms, soliton molecules, noise-like pulses, rogue waves

## 1 Introduction

Solitons[1,2] appear as localized structures in nonlinear systems, maintained by the robust balance between nonlinearity, dispersion, gain, and loss. The solitons are commonly known as fundamental modes in fluid mechanics[3], plasmas[4], nonlinear optics[5], Bose-Einstein condensates[6], etc. Leveraging the versatile nonlinear dissipative platform provided by mode-locked fiber lasers[7,8], transient dynamics of solitons was experimentally and theoretically studied in diverse settings by employing techniques such as time-stretch dispersive Fourier transform (TS-DFT)[9,10], and using theoretical models based on generalized nonlinear Schrödinger equations (GNLSEs)[11,12]. These include soliton explosions[13], rogue waves[14], soliton pulsations[15], multicolor solitons[16], breathers[17], talbot solitons[18], etc. The studies of the nonsteady soliton dynamics in these physical systems significantly contribute to the advancement of the fundamental nonlinear science and applications, such as optoelectronics.

The exploration of soliton buildup in fiber cavities began in the 1990s[19,20]. However, the real-time spectral buildup process was investigated by means of the TS-DFT technique[21] until 2010s. The latter work had revealed two scenarios of the soliton buildup in the negative-dispersion regime. The first scenario features four distinct stages, *viz*., the rising relaxation-oscillation (RO), quasimode-locking (QML), spectral-beating, and stable mode-locking ones. Another scenario features additional transient bound states, observed prior to the establishment of stable solitons. The formation of dissipative solitons in the normal-dispersion regime involves additional factors, such as gain and loss, being often accompanied by the generation of high-energy pulses[21-23]. The entire formation and evolution scenario for dissipative soliton features four stages: RO, spectral broadening, transient interaction, and the establishment of stable dissipative soliton states.

---

[*] Corresponding author email: dcq424@126.com

The buildup mechanism for "soliton molecules" (SMs, i.e., two-soliton bound states, that were predicted[24] and first experimentally reported[25]) has also drawn much interest, typically involving stages of mode locking, soliton splitting, and soliton interaction[26-28]. Besides, the special buildup scenarios such as the short-time and long-time SM formation[29] dynamics could be lead by gain/loss. In the former case, repulsive interactions dominate due to large differences in the initial pulse intensity and small initial separation between the interacting pulses. In contrast, the long-time scenario gives rise to phenomena such as soliton splitting, energy transfer between solitons, and attractive interactions between them. These buildup scenarios include stages that every soliton and SM must go through. SMs offer highly promising applications for the creation of efficient optical frequency combs[30] and multi-channel optical data encoding[31,32]. Exploring new soliton buildup mechanisms is significant for the creation and stabilization of SMs under specific conditions. The exploration of novel buildup mechanisms remains a relevant problem, which may help to optimize lasers and minimize the impact of extreme optical configurations.

After the formation of SMs was reported in many works, SMs have become a hot research topic – in particular, the unique particle-like mechanism of SM has become a research hotspot under the influence of self-phase modulation (SPM), cross-phase modulation (XPM), four-wave mixing, and group-velocity dispersion (GVD) and other factors. Benefitting from the technology of real-time spectral measurements, diverse isomers[33] of SMs and their temporal reconstruction processes[34] have been discovered. Furthermore, the existence of bound soliton complexes under the action of pure-high-even-order GVD (of orders 4-10, in the absence of the usual lower-order GVD[35]) has also been demonstrated. Additionally, the pump excitation may lead to the formation of soliton or SM clusters, which exhibit various dynamical effects, such as breathing dynamics in dual-color SMs[36] and synchronization of long-period oscillations in twin SMs[37]. Meanwhile, the nonlinear Fourier transform (NFT)[38,39] has also been applied, as a signal processing method, to control the formation, drift and collision between optical SMs, which enhances the analysis dimensionality in the transient states and further advances the control of the SM dynamics.

Different from SMs, noise-like pulses (NLPs)[40,41] with a exist in highly nonlinear cavity environments with a broad bandwidth. The NLP is related to soliton turbulence arising from strong interactions in broadband nonlinear fields, and reveals typical characteristics of the soliton turbulence with multiple ultrafast local structures. This fact helps to understand nonlinear phenomena including the soliton fission, Raman-driven evolution, and supercontinuum generation[14]. Furthermore splitting of NLPs gives rise to NLP complexes[42]. As features of highly nonlinear environments, NLPs serve as origins of extreme optical events (rogue waves)[43], helping to generate optical signals with extremely high amplitudes, that facilitate more effective signal detection. Additionally, precise switching between conventional solitons can be achieved by adjusting the cavity loss-and-gain balance, enhancing schemes for the data encoding[44]. These findings underscore the value of NLPs for fundamental studies and applications. However, the the switching mechanism between NLPs and solitons, SMs and soliton clusters has not yet been explore in necessary detail.

In this work, we discover a novel buildup mechanism of SMs involving long-duration NLPs, which breaks through the traditional understanding of the SM buildup and other previously studied nonlinear regimes. Employing the real-time spectral measurement technique and correlation testing method, we conduct a systematic analysis of the reversible polarization-controlled switching between SMs and NLPs in single pulse, pulse pairs, and pulse clusters under the action of strong nonlinearity and dispersion. The switching between pulses with different intensities is often accompanied by extreme soliton-turbulence events featuring complex interactions. In this work, the experimental study and validation are systematically conducted on the transmission patterns of SMs and NLPs. Further, we reproduce typical experimentally observed characteristics of the pulse-switching mechanism by simulations. Thus, the present work contributes to the understanding of the buildup and switching dynamics of highly complex nonlinear states, and offers innovation for the optical-data encoding and related topics.

## 2 The experimental device and results
### 2.1 The experimental setup

As the setup for this study, the mode-locked fiber laser (MLFL) is presented in Figure 1(a). The pump light, provided by a 980 nm laser diode, enters the ring cavity through a wavelength-division multiplexer (WDM). The pump light is amplified by a 10 m erbium-doped fiber (EDF) with dispersion 0.028 ps$^2$/km. Then, the light is transmitted through a 12.3 m single-mode fiber (SMF) with dispersion 0.0217 ps$^2$/km, which makes the net dispersion in the fiber equal to 0.017 ps$^2$. The 10m EDF enables the 1.5 μm light signal in the cavity to be fully excited by ions Er$^{3+}$. The saturable absorber (SA) used in this work is MOF-253@Au, which is a regular rod-like structure formed by the coordination bonding between Au$^+$ and ligands (C, N, O, Al). As indicated in the yellow box in the experimental setup, MOF-253@Au exhibits 30% saturation absorbance, 33.6% modulation depth, and a low saturation intensity of 0.039 MW/cm$^2$ at the wavelength of 1.5 μm. These advanced optical properties and the unique cavity design enables the laser output and soliton splitting easier in a highly nonlinear cavity environment, which is a prerequisite for the occurrence of NLPs and SM. Additionally, a 10-km SMF connected to a high-speed oscilloscope allows the observation of the novel buildup process of SMs and their real-time switching into NLPs. The spectrum analyzer, autocorrelator, and frequency analyzer are used to measure the average spectral characteristics, temporal characteristics, and frequency characteristics of SMs and NLPs, respectively.

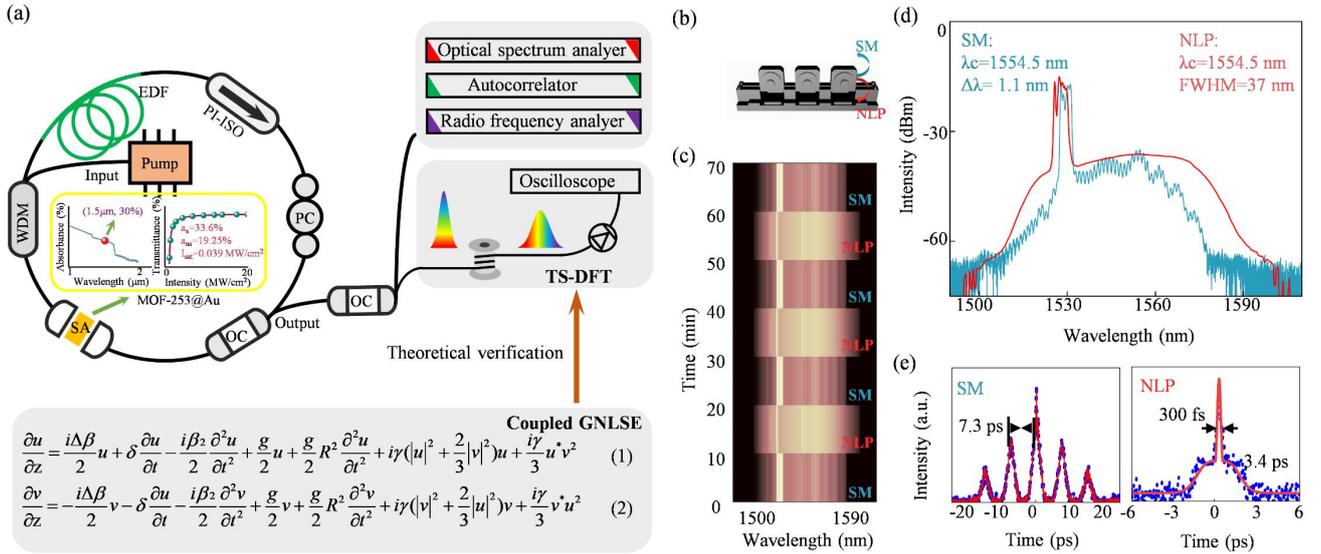

Figure 1 The experimental setup and results. (a) A schematic diagram of the setup; (b) the active control of intracavity losses; (c) the SM-NLP switching and stability under the action of different loss strengths; (d) average spectral profiles; (e) SM and NLP autocorrelation traces.

The theoretical model for the setup displayed in Figure 1(a) is based on coupled GNLSE:

$$\frac{\partial u}{\partial z} = \frac{i\Delta\beta}{2}u + \delta\frac{\partial u}{\partial t} - \frac{i\beta_2}{2}\frac{\partial^2 u}{\partial t^2} + \frac{g}{2}u + \frac{g}{2}R^2\frac{\partial^2 u}{\partial t^2} + i\gamma(|u|^2 + \frac{2}{3}|v|^2)u + \frac{i\gamma}{3}u^*v^2 \quad (1)$$

$$\frac{\partial v}{\partial z} = -\frac{i\Delta\beta}{2}v - \delta\frac{\partial u}{\partial t} - \frac{i\beta_2}{2}\frac{\partial^2 v}{\partial t^2} + \frac{g}{2}v + \frac{g}{2}R^2\frac{\partial^2 v}{\partial t^2} + i\gamma(|v|^2 + \frac{2}{3}|u|^2)v + \frac{i\gamma}{3}v^*u^2 \quad (2)$$

which are written for the amplitudes of optical waves $u$ and $v$ with perpendicular polarizations, $2\Delta\beta = 2\pi\Delta n/\lambda$, $2\delta = 2\Delta\beta\lambda/2\pi c$ and $\Delta n$ are the respective wavenumber, inverse-group-velocity, and refractive-index birefringence, respectively. Next, $\beta_2$ and $\gamma$ are the group-velocity dispersion (GVD) and Kerr-nonlinear coefficients. For the EDF, $R = 1/\Omega_g$ with gain bandwidth $\Omega_g$, and the gain saturation is modeled by

$$g = \frac{g_0}{(1+\int(|u|^2+|v|^2)dt/E_{\text{sat}})}, \quad (3)$$

where $g_0$ is gain coefficient for weak signals, and $E_{\text{sat}}$ is the gain-saturation energy related to the pump

power. The SA transmittance $T$ is

$$T = 1 - \alpha_{ns} - \frac{\alpha_s}{[1+(|u|^2+|v|^2)/I_{sat}]}, \qquad (4)$$

where $\alpha_{ns}$ is the unsaturated loss, $\alpha_s$ is the modulation depth, and $I_{sat}$ is the saturation power. In addition, the gain/loss caused by the intracavitary polarization controller (PC) is represented by the Jones matrix:

$$J_{pc} = \begin{bmatrix} \cos\theta & -\sin\theta \\ \sin\theta & \cos\theta \end{bmatrix} \begin{bmatrix} e^{\frac{i\varphi}{2}} & 0 \\ 0 & e^{-\frac{i\varphi}{2}} \end{bmatrix} \begin{bmatrix} \cos\theta & \sin\theta \\ -\sin\theta & \cos\theta \end{bmatrix}, \qquad (5)$$

where $\theta$ is the rotation angle of PC and $\varphi$ is the phase delay. The model incorporates the physical effects, such as the dispersion, nonlinearity and gain. It accurately simulates the pulse transmission in real fiber systems, especially the soliton transmission and switching under extreme conditions.

In this setup, the rotation angle of the PC in Figure 1(b) is set by means of the active control. The switching between SM and NLP in Figure 1(c) is realized, with each state being maintained for 10 min, indicating their excellent stability. The average spectral states of SM and NLP are shown in Figure 1(d). The central wavelength of SM and NLP is both 1554.5 nm, and the spectral fringe spacing $\Delta\lambda$ of SMs is 1.1 nm. NLPs have a typical spectral width 37 nm, which is wider than in the case of ordinary SMs. The obvious peak at 1530 nm in Figure 1(d) is due to the excessive energy in the cavity, which produces a continuous wave with a high peak value[45,46]. In Figure 1(e), the pulse time interval $\Delta t$ of the SM is 7.3 ps, which is consistent with the value calculated theoretically for parameters corresponding to the spectrum of the SM in Figure 1(d), viz., $\Delta t = |\lambda_c^2/(c \cdot \Delta\lambda)|$, where $\lambda_c$, $c$, $\Delta\lambda$, $\Delta t$ are central wavelength, speed of light, fringe interval, and time interval, respectively. The right side of Figure 1(e) shows the typical time state of the NLP, whose base- pulse width is 3.4 ps, and a very high-intensity spike with a pulse width of 300 fs is observed in the center. A wider spike means that the coherence between sub-pulses is high. The broadband smooth average spectral envelope and autocorrelation curve with a broad base and narrow peaks demonstrate the successful establishment of the NLP state[43,44,47]. These results prove that our objects are NLPs and SMs composed of three solitons.

## 2.2 Novel buildup process of SMs involving NLP

Benefiting from the strong nonlinear environment created by MOF-253@Au, the stable existence of SMs can be controlled externally by perturbations, viz., the gain/loss and polarization changes induced by PC and the pump power. With the input of 128 mW pump light, ample positive feedback is provided for the self-organization of SMs, which leads to two buildup scenarios: the usual one, and the scenario involving long-duration NLP, as shown in Figure 2(a). By pumping input into the fiber cavity, SM 1 undergoes several stages of the evolution, including noise, RO, soliton interactions and stable bound states, which is consistent with the previously reported mechanism[26].

In contrast to the traditional buildup scenario of SMs[26,29], the buildup process of SM 2 includes an NLP stage between the relaxation-oscillation and soliton-interaction ones. By means of the noise-driving screening, the specific frequency noise in the cavity is sufficiently stimulated by the positive feedback and enters the RO stage. The pulse energy of SM 2 at the RO stage is higher than that in the case of the usual buildup scenario (SM 1). Excessive SPM and the fission of noise pulses rapidly broaden the spectrum. The SM cannot withstand such a degree of spectral broadening, carrying the instability-inducing extremely high energy, therefore the chaotic envelope of the noise can only demonstrate a stage of transition towards SM, namely, the broadband spectral NLP, which transits to the soliton-interaction stage after the energy offsetting during 464 round trips (RTs). Comparing the energy in Figure 2(b) with the peak-power evolution trajectory in Figure 2(c), at time A (the 1309th RT), SM 2 is first screened out from the noise and enters the RO stage, then forms NLP and acquires the maximum energy to generate a wave with an extremely high amplitude. The high energy helps another temporally localized noise pulse to acquire the sufficient stimulation for starting RO at time B (the 1398th RT). After passing the stage of the soliton interaction, SM 1 is preliminarily formed. The soliton-interaction stage of SM 1 ends at time C (the 1862th RT), which coincides with the end of the NLP state of SM 2. During the interval between time B and time C, the high energy from the long-duration NLP in SM 2 generates the high-amplitude wave, and creates a complex highly nonlinear environment in the cavity, thus affecting SM 1 during the same period and producing random interaction forces within the SMs. This leads to spectral interference fringes during the interaction stage in Figure 2(d), and temporal oscillations during the self-correlation evolution in Figure 2(e), for which the autocorrelation evolution of the pulse is obtained by the first-order Fourier transform[48] of the spectrum in Figure 2(d). After time C, the adjustment effect of the dispersion compensates the modulation instability caused by the SPM during the B-C period, and gradually stabilizes the energy of SM 1. After the calculation of the corresponding function of the spectral modulation period and the time interval, and the time-stretching function[49] $\Delta\lambda_0 = \Delta t_0 / (|D| \cdot L)$, where $\Delta t_0$, $\Delta\lambda_0$, D, L are the duration of the stretched pulses, their spectral width, the dispersion parameter and the length of the dispersive medium, the time interval between the three solitons in the stable SM1 are 5.3 ps and 10.2 ps, respectively. The time intervals are basically consistent with the result produced by the first-order autocorrelation trajectory, which are 5.1 ps and 9.9 ps. However, the energy of SM 2 still exhibits slight fluctuations for 738 RTs (from the 1862th RT to 2600th RT) after time C, accompanied by fluctuations in the peak power in Figure 2(c). From the spectral interference fringes featured by the fluctuation stage in Figure 2(f), SM 1 is influenced by the extreme cavity environment during the long-duration NLP stage, experiencing excessive SPM until the dispersion adjustment is completed at the 2600th RT. The soliton motion trajectory in Figure 2(g) indicates that the interference effect of SM 2 from the 1862th RT to 2600th RT is due to the strong attractive and repulsive forces between solitons, and SM 2 stably propagates after the 2600th RT. Meanwhile, the soliton interval of SM 2 are 4.57 ps and 8.28 ps, matching the experiment.

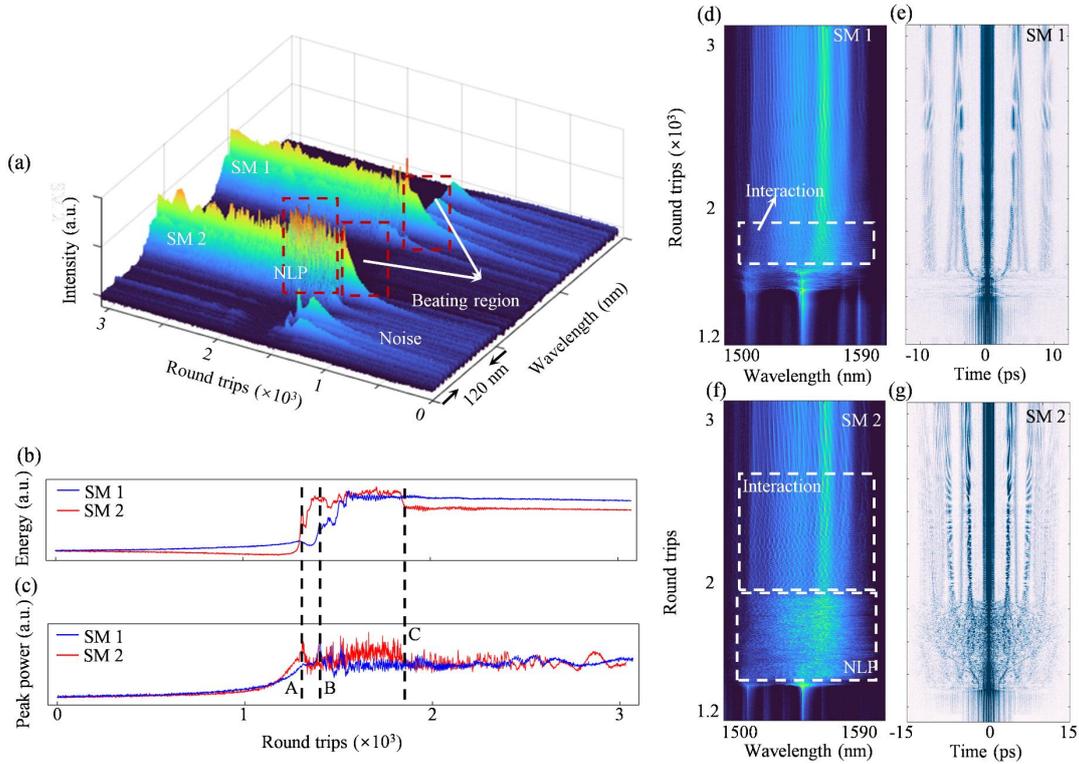

Figure 2. The SM buildup dynamics. (a) Buildup processes of the two SMs; (b) the energy and (c) peak power during the establishment process, (d), (f) the spectral and (e), (g) autocorrelation evolution of (d), (e) SM 1 and (f), (g) SM 2.

### 2.3 The Mechanism for Switching Between SMs and NLPs

To explore the correlation and switching mechanism between NLPs and SMs, and further achieve better control over the cavity state, we conduct a systematic analysis of switching between NLPs and SMs in the form of single-pulse, pulse-pair, and pulse-cluster states.

*The switching between SM and NLP.* Figure 3 depicts the complete transition between SM and NLP. In particular, Figures 3(a) and 3(b) present the spectral dynamics and their autocorrelation traces within 3000 RTs with the input of 128 mW pump light. By rotating the PC angle within a fixed range, the polarization state changes in the cavity. Increasing external disturbances, such as polarization losses, disrupt the balance of the nonlinearity and dispersion in the cavity, and thus the medium-scattering effect alters the optical frequency and phase of pulse. Simultaneously, SM cannot withstand the strong energy influx in the cavity, and the modulation effect of the dispersion fails to compensate the excessive SPM and higher-order nonlinear effects experienced by the pulse. As a result, the SM collapses and a new soliton state, namely NLP, establishes. Before collapsing from the 1280th RT to 1420th RT, SM undergoes a varying periodic intensity modulation process due to the irregular inter-soliton interaction in Figure 3(c). This is attributed to the imbalance among the medium scattering, SPM and dispersion. Figure 3(d) indicates that, before the 1280th RT, SM remains in a stable state with small energy oscillations and slight fluctuations of the peak power. Starting from the 1280th RT, the energy fluctuations of SM become a strong disturbance, which corresponds to the significant periodic perturbations of the spectral intensity and spectral density. Concurrently, the collapse of the SM state leads to a collapse in the energy and peak intensity, implying that an extremely unstable strong nonlinear environment appears in the cavity. At this time, NLP is established by the self-organization of the collapsing pulses. During this self-organization process, under the condition of the balance between SPM and dispersion, the pulse undergoes energy overshoot, and realizes the re-establishment of NLP by the energy exchange with the medium, which exhibits a typical characteristic of the shot-to-shot instability caused by the nonlinear phase shift due to the scattering effect of the medium. The switching stage demonstrates spectral oscillations, spectral collapse, and self-organization. Compared to previous works, these are essential features of the soliton switching in various nonlinear settings[34,50]. As reported above in Figure 2, NLP can exist as the transition state of SM during the buildup stage, thus SM and NLP can switch between each other in the strongly nonlinear environment, by setting appropriate cavity parameters.

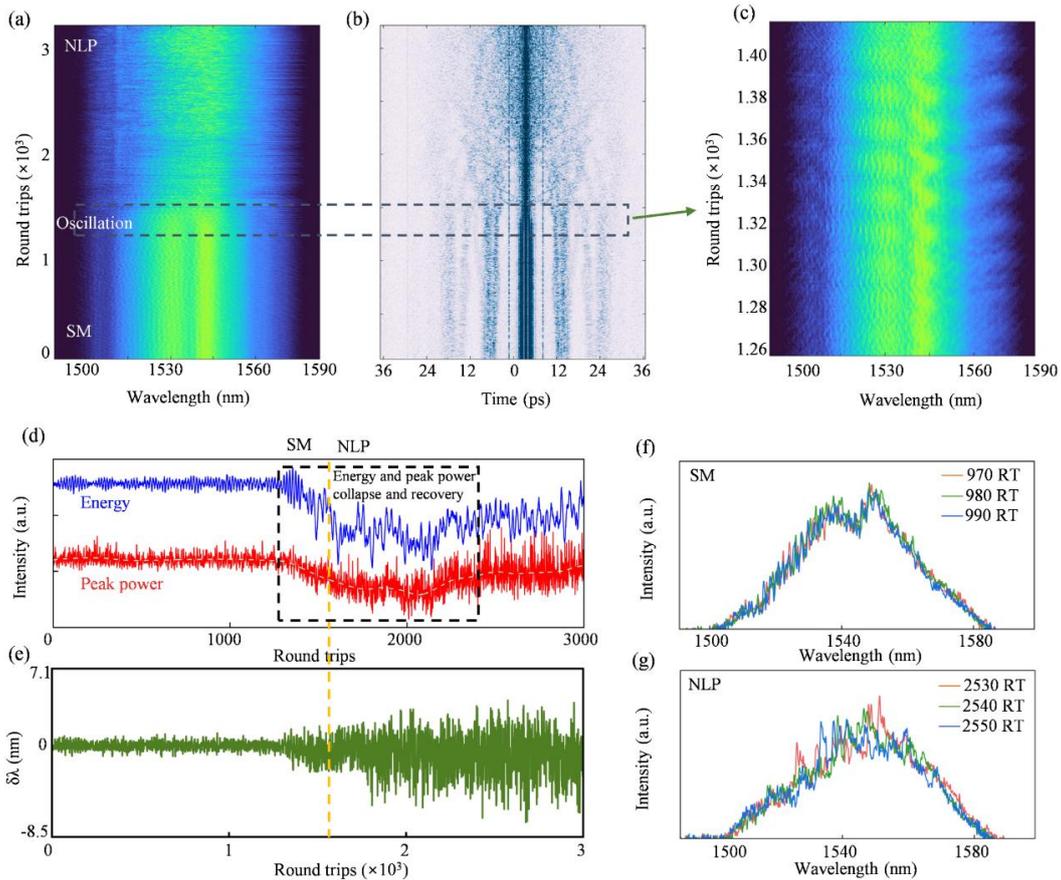

Figure 3 The dynamics of the SM-NLP switching. (a) The spectral and (b) autocorrelation evolution by DFT; (c) the spectrum from the 1260th RT to 1420th RTs; (d) the energy and peak power; (e) the real-time maximum wavelength offset within 3000 RTs, and spectral cross-sections of (f) SMs and (g) NLPs at three different RTs.

During the switching from SM to NLP in Figure 3(e), the shift of the maximum wavelength of the spectrum demonstrates that the spectral instability gradually develops before and after the state switching. In the stable stage, the maximum wavelength of SM remains steady within a small fluctuation range in Figure 3(f). At the switching stage, the fluctuation range of the maximum wavelength gradually increases, and the transition to the NLP state is achieved at the 2250th RT. Subsequently, the fluctuation range of the maximum wavelength of NLPs becomes random, which indicates the large-scale variation in the positions of the spectral maximum, as shown in Figure 3(g).

*The switching between SM and NLP pairs.* As mentioned above, the pulse collapse and subsequent self-organization are key features in the mechanism of the switching between SMs and NLPs in the broadband highly nonlinear cavity. In particular, the spectral collapse occurs the context of the soliton-SM interaction, while the self-organization of NLP is inseparable from the random splitting and interaction of solitons. In Figure 4, we elucidate the complete process of switching between SM and NLP pairs when the pump power is 130mW. Initially, by appropriately adjusting the PC to alter the cavity losses, Figures 4(a)-(d) depict the temporal state, spectral state, and autocorrelation evolution of the switching between the SM and NLP pairs. Pulse ① in Figure 4(a) comprises two pulses with a time interval of 3.3 ns, whose different frequencies are stretched and broadened in the highly dispersive medium to form a wideband SM spectrum through their interaction. Meanwhile, pulse ② on the right-hand side of Figure 4(a) is formed by a single wide pulse. Subsequently, the time-frequency relationship of the bound SMs in pulse ① and pulse ② remain constant with a time interval of 42.2 ns, due to the XPM interaction between pulses ① and ②. The peak-power curves for pulses ① and ② in Figure 4(e) reveal that pulse ① with the lower intensity is first affected by the medium scattering, and its peak power collapses at the 1321th RT, indicating the collapse of the SM state at this position. Meanwhile the pulse ② with high intensity exhibits a collapse delay of 60 RTs compared to pulse ①. These results imply that the higher intensity leads to superior robustness against the interference and superior robustness. After the collapse of pulse ① and pulse ②, the unstable positive gain remains in the cavity, and a self-organization process caused by the SPM and dispersion begins. The interaction between the two pulses caused by the XPM demonstrates a complementary trend in the peak-power curves of pulse ① and pulse ② in Figure 4(e). Using the Pearson correlation coefficient to assess the correlation between pulses ① and②, the correlation curve in Figure 4(f) shows that there is strong correlation, with the coefficient > 0.8, at the SM-pair stage of the evolution. During the switching from the SM pair to the NLP one, the correlation coefficient decreases and experiences significant fluctuations, which implies that the XPM effect strives to maintain the dynamical connection between the pulse pairs in terms of the intensity, energy and evolution, as they undergo the decoherence. In the NLP-pair state, the correlation coefficient between pulses ① and② is significantly lower than that in the SM-pair state. However, as the peak intensities of both pulses approach each other, the correlation coefficient of the NLP pair exhibits a peak again, which indicates that the closer their energies, the more pronounced is the modulation effect of XPM, and the more similar are their spectral states and temporal trajectories.

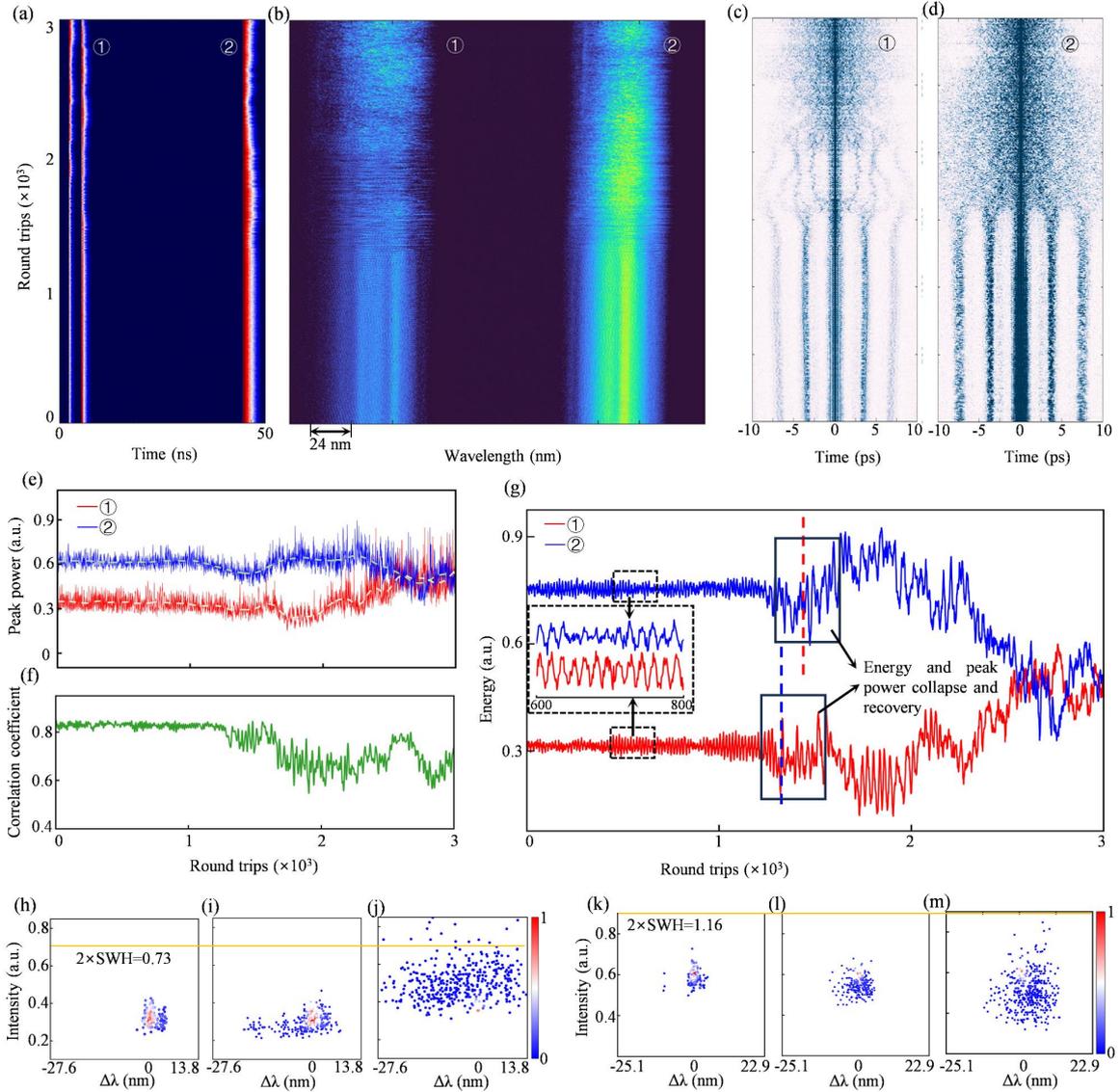

Figure 4 The switching dynamics between the SM and NLP pairs. (a) Temporal, (b) spectral, and (c,d) autocorrelation cross sections of pulses ① and②; (e) the peak power; (f) the correlation coefficient; (g) energies of pulses ① and②; intensity-distribution density maps of (h)-(j) pulse ① and (k)-(m) pulse ② at SM, switching, and NLP stages, respectively. Here the yellow lines represent their positions at 2×SWH.

Furthermore, the energy curves of pulses ① and② in Figure 4(g) reveal dependences similar to those for the peak power during the switching process, accompanied by typical characteristics observed during the SM-to-NLP switching in Figure 3. It includes the energy-instability, energy-collapse, and energy- overshoot self-organized behavior. The trajectory change of the peak power in Figure 4(e) and the change of the correlation coefficient in Figure 4(f) are mapped to the energy exchange behavior: the energy difference between pulses ① and ② attains a maximum at around 1850th RT, which incurs from the excessive SPM generated by the medium scattering in the course of the reconstruction of their states. Subsequently, as the nonlinear effect in the pulses diminishes, the XPM interaction between the pulses gradually becomes the factor which dominates the energy exchange, which leads to the convergence of the energy flow efficiency between two NLPs and the medium, and thus the overlapping of the energy intensity occurs from the 2550th RT to 2730th RT. At this position, the nonlinearity and dispersion in in the two NLPs are relatively balanced, meaning that the evolution of the NLP pair has reached a state of the mutual balance, and they adhere to the energy complementarity principle.

Finally, the relationships between the maximum wavelength and intensity of the SM, switching, and NLP state during the spectral evolution of pulses ① and ② are characterized in Figures 4(h)-(j) and Figures 4(l)-(n), respectively. This plots demonstrate the intensity distribution patterns in the different states: the intensities and maximum wavelengths of pulses ① and ② are relatively concentrated during the SM state, diverge within a certain range during the switching state, and achieve complete randomness in their positions during the NLP state. By calculating the significant wave height (SWH) of pulses ① and ②, it is evident that extremely high-amplitude rogue waves, exceeding 2×SWH, appear in pulse ① but not in pulse ②. This indicates that the wave packet with low energy is more susceptible to the disruption of its original envelope by the highly nonlinear cavity perturbation, while the wave packet with high energy can maintain better stability when facing extreme environments.

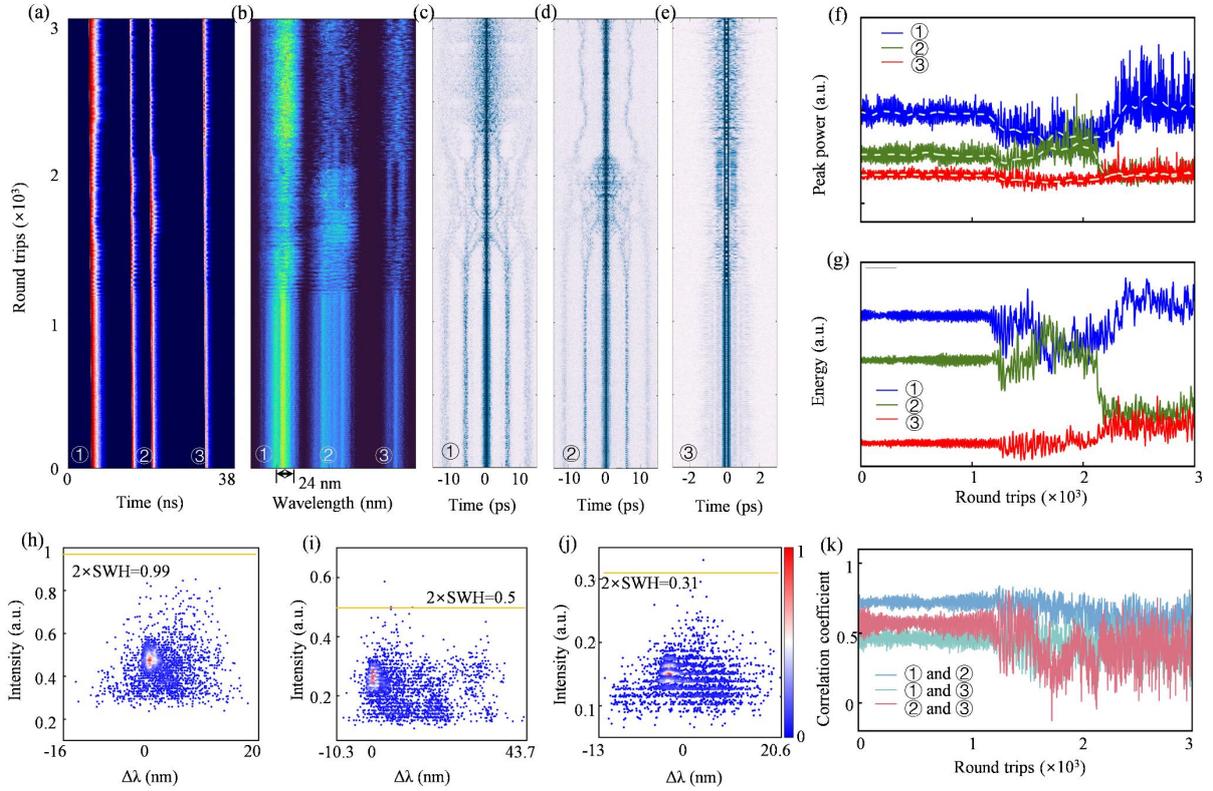

Figure 5. The switching dynamics between the SM and NLP clusters. (a) The temporal evolution over 3000 RTs; (b) the spectral evolution in the cluster. Autocorrelation evolution traces for pulses pulse ① (c), pulse ② (d); and pulse ③ (e). (f) The peak power and (g) energy curves for pulses ①, ②, and ③; intensity-distribution density for pulses pulse ① (h), pulse ② (i) and pulse ③ (j). The yellow line denotes the position of 2×SWH. (k) The correlation index between the pulses.

*Switching between SM and NLP clusters.* To broaden our understanding of the switching of SMs and NLPs under various conditions and their interconnected switching mechanisms, we here analyze the switching of three-pulse clusters, and elucidate the full process of switching between SM and NLP clusters, as shown in Figure 5(a). The temporal evolution of this switching scenario is similar to the temporal characteristics of the SM pair in Figure 4(a), namely, there also exist two nearby pulse pairs in the SM cluster, and they undergo broadening due to the nonlinear phase shift of pulses, forming SM, as illustrated by pulse ② in Figure 5(b). The time-spectral mapping between three distant pulses ① ② ③ in Figure 5(b) indicates that the XPM in the cavity dynamically locks the inner relationship in the-pulse cluster. After stable evolution of the SM cluster about 1150 RTs, a sudden change in the gain/loss in the cavity, induced by the actively controlled polarization, disrupts the internal balance of the SPM and dispersion effects in each pulse, which results in a similar fixed feature, just like demonstrated by the SM and SM pair before the switching stage, *viz.*, spectral oscillations with a variable period. The autocorrelation traces of pulses ①, ②, and ③ during the switching process in Figures 5(c)-5(e) demonstrate that each pulse undergoes the soliton splitting and instantaneous interaction in the course of the state switching. When the energy carried by the pulse exceeds its own limit, the splitting occurs, which is also the same effect as exhibited by NLP. From the autocorrelation trace of pulse ② in Figure 5(d), we observe the switching between SM and NLP complexes, that is, after the SM passes the soliton-interaction stage, a pattern carrying NLP complexes is established in the cavity. The peak power in Figure 5(f) and energy in Figure 5(g) of pulses ①, ②, and ③ reveal that XPM maintains the energy exchange between the medium and pulses, and stabilizes interactions between pulses during their state transitions.

Tracking the energy and peak power of pulse ②, we find that, during the collapse and subsequent energy rebound of pulse ② from the 1220 RT to the 1700 RT, pulse ① close to pulse ② is most affected, exhibiting significant intensity fluctuations. In contrast, pulse ③ experiences less influence from XPM and results in minor fluctuations of the energy exchange, due to its greater distance from pulse ②. At the 2122th RT, the energy and peak power of pulse ② suddenly exhibit obvious loss. At the same time, the lost energy is absorbed by the medium and transferred by the energy exchange with pulses ① and ③, so that these pulses receive additional energy. This phenomenon occurs because the inability of NLP complexes to carry excessively high energy leads to energy outflow, which is absorbed by NLP at the same time. It originates comes from the principle of equilibrium between the dispersion and strong nonlinearity, and again adheres to the above-mentioned energy complementarity principle in the cavity. Furthermore, the intensity-distribution (density maps) of pulses ①, ②, and ③ clearly illustrate the relationship between intensity distributions during the SM to NLP switching in Figures 5(h)-5(j). No rogue wave appears in pulse ① with higher intensity during the switching process, whereas it is detected in both weaker pulses ② and ③, which implies that the high-intensity NLP exhibits higher stability in the turbulent environment with strong interactions, while weaker pulses are more vulnerable to disturbances in the cavity, and thus increase the probability of extreme events.

The spectral correlation between pulses ①, ② and ③ is assessed by means of the Pearson correlation coefficient[43] in Figure 5(k). Due to the time interval of pulses in the cluster, in the SM states the correlations between pulses ① and ②, pulses ② and ③, and pulses ① and ③ are attenuated successively. However, in the NLP stage, the correlation between the pulses is less sensitive to the time interval between them. In summary, the switching mechanism between SM and NLP for the multiple-pulse sets is similar to that for the single pulse in the cavity. However, considering the non-negligible XPM of multiple pulses, the energy transfer occurs in the switching of multiple pulses, which is crucial to maintain the stability of the cavity dynamics.

## 3 Theoretical Simulations
### 3.1 Verification of novel buildup process of SM involving NLP

Based on the experimental setup described in Figure 1, the round-trip model is established. Then buildup and switching mechanisms of SM involving NLP is verified by simulations of the system of coupled GNLSEs (1) and (2) with parameters collected in Table 1. The simulations were performed by means of the split-step Fourier method, which provides excellent computational accuracy and efficiency for this purpose. In the simulations, we employ a higher Kerr-nonlinearity coefficient than the one used for general simulations which aim to reach the special critical point for the SM splitting in the cavity with accumulated excessive nonlinearity. It allows the appearance of long-duration NLP state alongside the establishment of SM. Figure 6(a) depicts the complete temporal evolution process during the SM buildup stage, including the noise, NLP, soliton interaction, and stable bound states, which thereby validates the buildup process of SM involving long-duration NLPs in Figure 2(a). From the right side of Figure 6(a), the peak-power fluctuations are larger at the early NLP stage than at the later SM one, which is consistent with the intensity descriptions of NLP and SM in Figure 2(c). From the autocorrelation profiles produced during the SM buildup stage at the 250th RT and 2000th RT in Figure 6(b), SM exhibits a typical multi-peak structure, while NLP features a central high peak and wide base. The change of spectral states in Figure 6(c) illustrates the transition from NLP to SM in Figure 6(a). The pattern of maximum wavelength offset in the right-hand side of Figure 6(c) demonstrates the disorderliness in the NLP state, while it is confined to a minimal range in the SM state, which is consistent with the spectral-shift characteristics demonstrated in Figure 3(e). Figure 6(d) further validates the authenticity of the NLP state during the SM buildup process, by contrasting the irregular and regular spectral envelopes of solid line in the NLP and SM states, respectively, and the smooth broadband spectrum depicted by the dashed line (the upper plot in Figure 6(d)) which represents the average state of NLP at 1000 RTs, and has the same characteristics as produced by the spectrum analyzer in Figure 1(d).

Table 1 Model parameters.

| Fiber parameters | SMF$_1$ Value | EDF Value | SMF$_2$ Value | SA | Value |
|---|---|---|---|---|---|
| Z (m) | 2 | 10 | 10.3 | $\alpha_{ns}$ | 0.336 |
| $\beta_2$ (fs$^2$/mm) | -21.7 | 28 | -21.7 | $\alpha_s$ | 0.193 |
| $\gamma$ (W$^{-1}$/m$^{-1}$) | 0.008 | 0.0036 | 0.008 | $I_{sat}$ (W/cm$^2$) | 0.039 |
| $\Delta\beta$ (m$^{-1}$) | 4.05 | 4.05 | 4.05 | PC parameters (rad) | Value |
| $\Omega_g$ (nm) | - | 30 | - | $\theta$ | 1.6 (buildup) 2.21 (switching) |
| $g_0$ (m$^{-1}$) | 0 | 12 | 0 | | |
| $E_{sat}$ (pJ) | - | 5 (buildup) 3 (switching) | - | $\varphi$ | 2.5 |

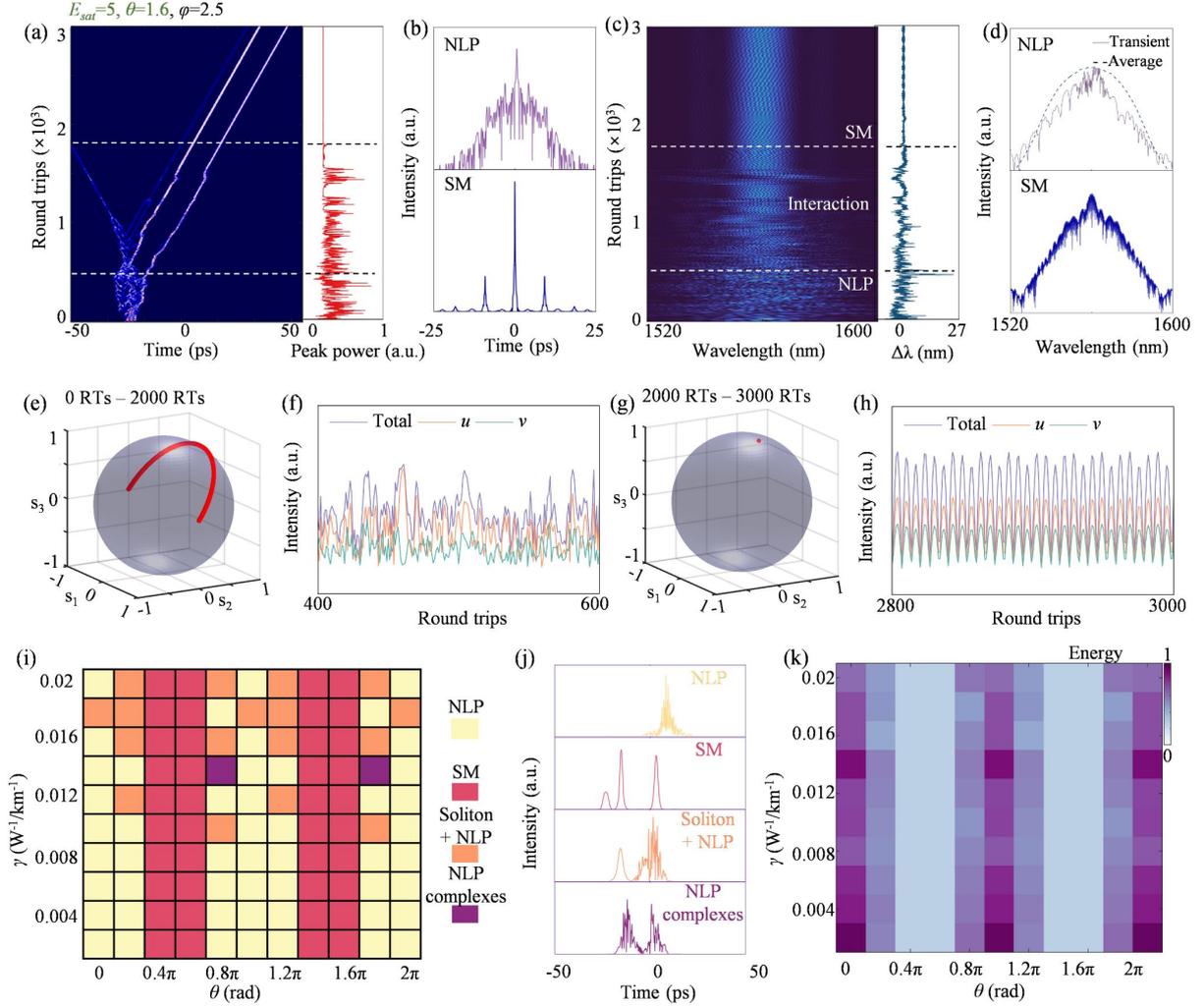

Figure 6. The numerically simulated SM buildup dynamics and systematic switching following the variation of $\theta$ and $\gamma$ in Eqs. (1), (2) and (5). (a) The temporal and (c) spectral evolution snapshots within 3000 RTs along with the corresponding peak-power curve and peak-wavelength offset, respectively. (b) Autocorrelation and (d) spectral cross-sections of NLP at the 250th RT and SM at the 2000th RT. The Stokes distribution during the buildup in intervals 0-2000 RTs (e) and 2000-3000 RTs (g). The total energy of components $u$ and $v$ are shown in panels (f) and (h), in intervals 400-600 RTs and 2800-3000 RTs, respectively. (i) The phase chart of the soliton switching in the steady state. (j) Typical time snapshots and (k) energy intensity of different soliton states corresponding to (i).

Subsequently, using the Stokes vector on the Poincaré sphere, we analyze the relation between the fast- and slow-axis vector components of the coupled system. Here the four standard Stokes parameters $s_j$ ($j$=0~3) are defined as $s_0 = |u|^2 + |v|^2$, $s_1 = |u|^2 - |v|^2$, $s_2 = 2|u||v|\cos(\Delta\Phi)$, $s_3 = 2|u||v|\sin(\Delta\Phi)$, where $\Delta\Phi$ represents the phase difference between the two orthogonal components. Because the value of $s_0$ does not change with time and space, the Stokes vector with components $s_1$, $s_2$, and $s_3$ moves on the sphere of radius of $s_0$. Figure 6(e) demonstrate that, in the course of the SM buildup stage, 0–2000 RTs, the polarization components constitute a polarization-rotating vector state, which is confirmed by the energy complementarity between the *u* and *v* components in the interval of 400–600 RTs in Figure 6(f). Conversely, Figure 6(g) illustrates that, within the interval of 2000-3000 RTs, the two components are tuning to a polarization-locked vector state, which corresponds to the synchronized energy oscillations in Figure 6(h).

The variation of the nonlinear parameter $\gamma$ in Eqs. (1), (2) and PC rotation angle $\theta$ in Eq. (5) leads to changes in the gain and loss acting in the cavity, thereby a phase transition in the soliton state occurs in the cavity. Figure 6(i) shows the chart of the soliton states corresponding to 3000 RTs, with $\theta \in [0, 2\pi]$ and $\gamma \in [0.002, 0.02]$. In the chart, the soliton exhibits the periodic behavior (with the period of $\pi$), with the variation of $\theta$. From Figure 6(i), when $\gamma \in [0.002, 0.008]$, varying the PC rotation angle $\theta$ in the Jones matrix between 0 and $0.4\pi$, $0.6\pi$ and $1.4\pi$, and $1.6\pi$ and $2\pi$, the pulse is in the NLP state, while between $0.4\pi$ and $0.6\pi$, and between $1.4\pi$ and $1.6\pi$ it is in the SM state. These results validate the experimental observations of the NLP-SM switching presented in Figure 1(c), which is induced by the interplay of the nonlinearity and polarization. At $\gamma > 0.008$, the excess nonlinearity leads to complex interactions in the cavity, which lead to the coexistence of solitons and NLPs, as well as special soliton states such as NLP complexes shown in Figure 6(j). The energy of map displayed in Figure 6(k) demonstrates that, as $\gamma$ varies, the internal energy of SM remains relatively stable, while NLP carries larger energy than SM.

**3.2 Numerical verification of the SM-NLP switching mechanism**

Further setting the parameters to $E_{sat} = 3$, $\theta = 2.21$, $\varphi = 2.5$, and using the 3000th RT from Figure 6(a) as the initial pulse, in Figure 7 we demonstrate results of the *switching* from SMs to NLP.

Figure 7(a) illustrates the time evolution of the switching from SM to NLP, where the strong attraction and weak repulsion lead to interactions between solitons in the SM. After that a complete transition from SM to NLP occurs at 1572th RT. The trend of peak power on the right-hand side indicates that an intensity collapse at the critical point of the SM-NLP transition is followed by rapid self-organization including the energy overshoot. These results validate the typical experimentally observed SM-NLP transition process from SM to NLP process presented in Figure 3. The spectral behavior during the switching process is elucidated in Figure 7(b). It shows that the wavelength offset pattern on the right-hand side is aligned with experimental observations of small-range fluctuations of the peak wavelength offset during the SM stage, and disordered fluctuations during the NLP one. Considering the strong correlation between the temporal and frequency structures of the wave packets, we analyze their time-frequency characteristics. using the short-time Fourier transform for both SM and NLP. Due to the combined effects of the dispersion and nonlinearity, as well as the interference between the temporal and spectral structures, multiple SM and NLP spots are observed in Figures 7(c) and 7(d), respectively. SM are distributed around the central frequency and correspond to the temporal positions of three solitons, while NLP is distributed symmetrically around the central frequency and is difficult to discern, which indicates the randomness of NLP evolution. Figure 7(e) demonstrate that the two vector components during the SM-NLP switching stage form a synchronous vector state. The two components during this SM-NLP stage correspond to a fixed point on the Poincaré sphere, which means that the rotating-polarization state is formed at the SM and NLP stages. The corresponding synchronous changes in the energy of NLP are shown in Figure 7(f).

Finally, the intra-cavity evolution of SM and NLP within 3 RTs was addressed. The simulations of both SM and NLP displayed in Figures 7(g) and 7(h) demonstrate that the negative dispersion of SMF and the positive dispersion of EDF give rise to opposite group velocities. It means that SMF and EDF cause convergence and divergence of the pulse envelope, respectively. The SM's energy attains extremely high values at specific positions in the second segment of SMF. In contrast, for the NLP, although the peak- intensity configuration occurs in the second segment of SMF, the position of the peak intensity lacks the regularity. In the spectral domain, the intra-cavity spectral envelope of the SM is excited by EDF, resulting in an increase of the spectral intensity in Figure 7(i). Conversely, due to the random distribution of frequency features observed in the time-frequency domain, the growth of the

NLP's intensity also exhibits the randomness when the NLP is amplified by EDF in Figure 7(j). In Figure 7(k), the evolution of the peak power, energy and spectral peak power is compared for the SM and NLP. For the SM, the peak intensity, energy and spectral intensity, which are amplified at the EDF position, are subsequently attenuated by the negative-dispersion effect of SMF, which eventually leads to the establishment of the relatively stable SM. However, for NLP, despite acquiring the intensity enhancement from the same EDF, the sudden action of the strong nonlinearity while passing the SMF causes the inability of the negative dispersion to balance the frequency offset which it induces, resulting in the final output of NLPs exhibiting the stochastic intensity distribution in the time-frequency domain. Due to the influence of the gain and loss, higher-order dispersion effects and special nonlinear effects, such as the Raman scattering, the GNLSE model cannot fully model the experimental setup[36,50]. Nevertheless, the present work predicts transient states that agree well with the experiment, barring minor details.

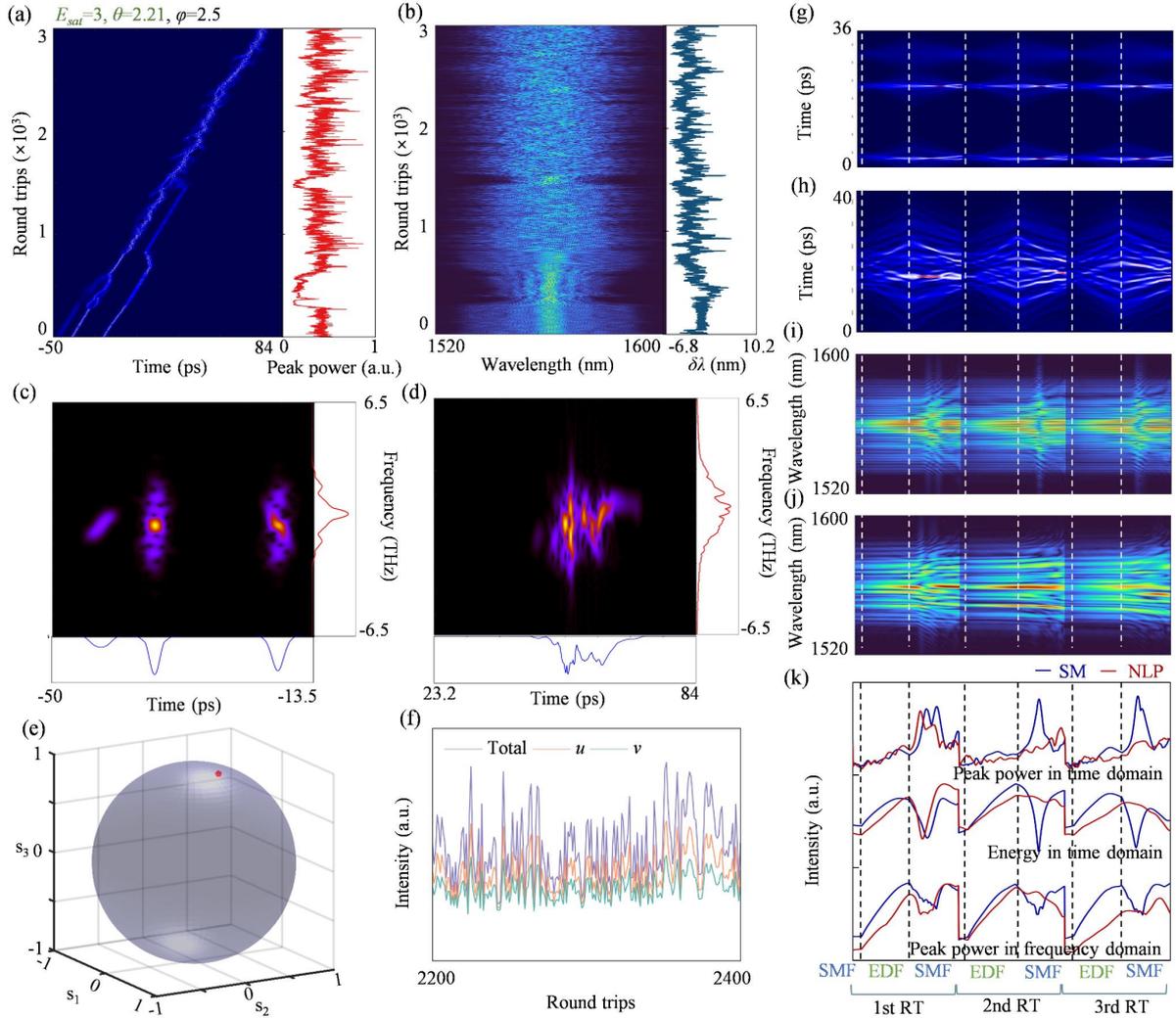

Figure 7. The numerically simulated switching dynamics between SMs and NLP. Snapshots of the he temporal (a) and spectral (b) evolution 3000 RTs. Time-frequency characteristics of (c) SM and (d) NLP states. The Stokes distribution during the switching process of SMs (f). The temporal and spectral evolution within 3 RTs for (g,i) SM and (h,j) NLP. (k) The temporal peak power, energy and spectral peak power curves for the SM and NLP states.

## 4 Conclusion

In summary, this study elucidates the novel scenario for the SM buildup, distinct from the usual one. The new scenario includes the stages of RO, formation of the long-duration NLP, soliton interactions, and formation of stable bound states. The scenario is experimentally demonstrated in the bimodal optical setup employing MOF-253@Au as SA and the polarization control by PC. The emergence of the NLP transition states offsets the strong interaction caused by the soliton turbulence and extreme nonlinearity in the cavity. The spectral collapse following spectral oscillation with a variable period, and the self-organization following the pulse's energy overshoots are found to be typical features of the SM-NLP switching, which occurs even if simultaneous multi-pulse switching takes place in the same temporal domain. In the scenarios involving the multi-pulse switching induced by the XPM-induced soliton interactions, the pulses demonstrate the specific energy-complementarity behavior. It means that the efficiency of the energy transfer between the pulses is affected by the temporal distance between them, *viz.*, as usual, the stronger energy transfer occurs at smaller distances. The intensity of the NLP state is positively correlated with its ability to maintain the stability. The low-intensity NLPs with poor stability lead to extreme optical events, such as the occurrence of rogue waves. Numerical simulations performed in this work reproduce the novel SM buildup mechanism involving the NLP transition state, along with typical features of the SM – NLP switching process. At specific values of the PC (polarization-controller) angles, the simulations demonstrate that phase transitions between SM and NLP states periodically occur in the nonlinear cavity, and produce other mixed soliton states observe din the experiment, validating the regularity and robustness of the soliton- switching regimes observed in the experiment. These findings provide a relevant addition to the traditional nonlinear soliton dynamics, and offer applications to the design of controllable multi-state laser sources.

**Funding**

National Natural Science Foundation of China (Grant Nos. 12261131495); the Scientific Research and Development Fund of Zhejiang A&F University (Grant No. 2021FR0009).

**Reference**

1 Kibler, B., Fatome, J., Finot, C., Millot, G., Dias, F., Genty, G., Akhmediev, N. & Dudley, J. M. The Peregrine soliton in nonlinear fibre optics. *Nature Physics* **6**, 790-795 (2010).

2 Opačak, N., Kazakov, D., Columbo, L. L., Beiser, M., Letsou, T. P., Pilat, F., Brambilla, M., Prati, F., Piccardo, M., Capasso, F. & Schwarz, B. Nozaki–Bekki solitons in semiconductor lasers. *Nature* **625**, 685-690 (2024).

3 Tomboulides, A. G., Triantafyllou, G. S. & Karniadakis, G. E. A new mechanism of period doubling in free shear flows. *Physics of Fluids A: Fluid Dynamics* **4**, 1329-1332 (1992).

4 Blyakhman, L., Gromov, E., Malomed, B. & Tyutin, V. Soliton oscillations in the Zakharov-type system at arbitrary nonlinearity-dispersion ratio. *Chaos, Solitons & Fractals* **117**, 264-268 (2018).

5 Peng, J., Zhao, Z., Boscolo, S., Finot, C., Sugavanam, S., Churkin, D. V. & Zeng, H. Breather Molecular Complexes in a Passively Mode-Locked Fiber Laser. *Laser & Photonics Reviews* **15**, 2000132 (2021).

6 Sakaguchi, H. & Malomed, B. A. Flipping-shuttle oscillations of bright one- and two-dimensional solitons in spin-orbit-coupled Bose-Einstein condensates with Rabi mixing. *Physical Review A* **96**, 043620 (2017).

7 Maniewski, P., Brunzell, M., Barrett, L., Harvey, C. M., Pasiskevicius, V. & Laurell, F. Er-doped silica fiber laser made by powder-based additive manufacturing. *Optica* **10**, 1280-1286 (2023).

8 Lang, J. A., Hutter, S. R., Leitenstorfer, A. & Herink, G. Controlling intracavity dual-comb soliton motion in a single-fiber laser. *Science Advances* **10**, eadk2290 (2024).

9 Zhang, H., Mao, D., Du, Y., Zeng, C., Sun, Z. & Zhao, J. Heteronuclear multicolor soliton compounds induced by convex-concave phase in fiber lasers. *Communications Physics* **6**, 191 (2023).

10 Li, Z., Liu, H., Zha, Z., Su, L., Shum, P. P. & Guo, H. Universal dynamics and deterministic motion control of decoherently seeded temporal dissipative solitons via spectral filtering effect. *Photonics Research* **11**, 2011-2019 (2023).

11 Fang, Y., Han, H.-B., Bo, W.-B., Liu, W., Wang, B.-H., Wang, Y.-Y. & Dai, C.-Q. Deep neural network for modeling soliton dynamics in the mode-locked laser. *Opt. Lett.* **48**, 779-782 (2023).

12 Wang, R.-R., Bo, W.-B., Han, H.-B., Dai, C.-Q. & Wang, Y.-Y. Vector pulsating solitons and soliton molecules under higher-order effects in passively mode-locked fiber lasers. *Chaos, Solitons & Fractals* **171**, 113438 (2023).

13 Si, Z.-Z., Wang, Y.-Y. & Dai, C.-Q. Switching, explosion, and chaos of multi-wavelength soliton states in ultrafast fiber lasers. *Science China Physics, Mechanics & Astronomy* **67**, 274211 (2024).

14 Meng, F., Lapre, C., Billet, C., Sylvestre, T., Merolla, J.-M., Finot, C., Turitsyn, S. K., Genty, G. & Dudley, J. M.

|  | Intracavity incoherent supercontinuum dynamics and rogue waves in a broadband dissipative soliton laser. *Nature Communications* **12**, 5567 (2021). |
|---|---|
| 15 | Wang, Z., Coillet, A., Hamdi, S., Zhang, Z. & Grelu, P. Spectral Pulsations of Dissipative Solitons in Ultrafast Fiber Lasers: Period Doubling and Beyond. *Laser & Photonics Reviews* **17**, 2200298 (2023). |
| 16 | Cui, Y., Yao, X., Hao, X., Yang, Q., Chen, D., Zhang, Y., Liu, X., Sun, Z. & Malomed, B. A. Dichromatic Soliton-Molecule Compounds in Mode-Locked Fiber Lasers. *Laser & Photonics Reviews* **n/a**, 2300471 (2024). |
| 17 | Wu, X., Peng, J., Boscolo, S., Finot, C. & Zeng, H. Synchronization, Desynchronization, and Intermediate Regime of Breathing Solitons and Soliton Molecules in a Laser Cavity. *Physical Review Letters* **131**, 263802 (2023). |
| 18 | Zhang, H., Du, Y., Zeng, C., Sun, Z., Zhang, Y., Zhao, J. & Mao, D. The dissipative Talbot soliton fiber laser. *Science Advances* **10**, eadl2125 (2024). |
| 19 | Pu, N.-W., Shieh, J.-M., Lai, Y. & Pan, C.-L. Starting dynamics of a cw passively mode-locked picosecond Ti:sapphire/DDI laser. *Opt. Lett.* **20**, 163-165 (1995). |
| 20 | Li, H., Ouzounov, D. G. & Wise, F. W. Starting dynamics of dissipative-soliton fiber laser. *Opt. Lett.* **35**, 2403-2405 (2010). |
| 21 | Peng, J. & Zeng, H. Build-Up of Dissipative Optical Soliton Molecules via Diverse Soliton Interactions. *Laser & Photonics Reviews* **12**, 1800009 (2018). |
| 22 | Zhang, Z., Tian, J., Cui, Y., Wu, Y. & Song, Y. Dynamics of multi-state in a simplified mode-locked Yb-doped fiber laser. *Chinese Optics Letters* **20**, 081402 (2022). |
| 23 | Zhao, H., Ma, G.-M., Li, X.-Y., Li, T.-J., Cui, H., Liu, M., Luo, A.-P., Luo, Z.-C. & Xu, W.-C. Buildup dynamics in an all-polarization-maintaining Yb-doped fiber laser mode-locked by nonlinear polarization evolution. *Optics Express* **28**, 24550-24559 (2020). |
| 24 | Malomed, B. A. in *Large Scale Structures in Nonlinear Physics.* (eds Jean-Daniel Fournier & Pierre-Louis Sulem) 288-294 (Springer Berlin Heidelberg). |
| 25 | Tang, D. Y., Man, W. S., Tam, H. Y. & Drummond, P. D. Observation of bound states of solitons in a passively mode-locked fiber laser. *Physical Review A* **64**, 033814 (2001). |
| 26 | Liu, X., Yao, X. & Cui, Y. Real-Time Observation of the Buildup of Soliton Molecules. *Physical Review Letters* **121**, 023905 (2018). |
| 27 | Song, Y., Shi, X., Wu, C., Tang, D. & Zhang, H. Recent progress of study on optical solitons in fiber lasers. *Applied Physics Reviews* **6**, 021313 (2019). |
| 28 | Liu, S., Cui, Y., Karimi, E. & Malomed, B. A. On-demand harnessing of photonic soliton molecules. *Optica* **9**, 240-250 (2022). |
| 29 | Zhou, Y., Ren, Y.-X., Shi, J., Mao, H. & Wong, K. K. Y. Buildup and dissociation dynamics of dissipative optical soliton molecules. *Optica* **7**, 965-972 (2020). |
| 30 | Weng, W., Bouchand, R., Lucas, E., Obrzud, E., Herr, T. & Kippenberg, T. J. Heteronuclear soliton molecules in optical microresonators. *Nature Communications* **11**, 2402 (2020). |
| 31 | Liu, Y., Huang, S., Li, Z., Liu, H., Sun, Y., Xia, R., Yan, L., Luo, Y., Liu, H., Xu, G., Sun, Q., Tang, X. & Shum, P. P. Phase-tailored assembly and encoding of dissipative soliton molecules. *Light: Science & Applications* **12**, 123 (2023). |
| 32 | Yang, Y., Lin, W., Guo, Y., Hu, X., Xu, H., Chen, D., Wei, X. & Yang, Z. Phase-encoding of loosely bound soliton molecules. *APL Photonics* **9**, 031305 (2024). |
| 33 | Huang, S., Liu, Y., Liu, H., Sun, Y., Xia, R., Ni, W., Luo, Y., Yan, L., Liu, H., Sun, Q., Shum, P. P. & Tang, X. Isomeric dynamics of multi-soliton molecules in passively mode-locked fiber lasers. *APL Photonics* **8**, 036105 (2023). |
| 34 | Zhou, Y., Shi, J., Ren, Y.-X. & Wong, K. K. Y. Reconfigurable dynamics of optical soliton molecular complexes in an ultrafast thulium fiber laser. *Communications Physics* **5**, 302 (2022). |
| 35 | Han, Y., Gao, B., Wen, H., Ma, C., Huo, J., Li, Y., Zhou, L., Li, Q., Wu, G. & Liu, L. Pure-high-even-order dispersion bound solitons complexes in ultra-fast fiber lasers. *Light: Science & Applications* **13**, 101 (2024). |

36  Cui, Y., Zhang, Y., Huang, L., Zhang, A., Liu, Z., Kuang, C., Tao, C., Chen, D., Liu, X. & Malomed, B. A. Dichromatic ``Breather Molecules'' in a Mode-Locked Fiber Laser. *Physical Review Letters* **130**, 153801 (2023).

37  Hamdi, S., Coillet, A., Cluzel, B., Grelu, P. & Colman, P. Superlocalization Reveals Long-Range Synchronization of Vibrating Soliton Molecules. *Physical Review Letters* **128**, 213902 (2022).

38  Sugavanam, S., Kopae, M. K., Peng, J., Prilepsky, J. E. & Turitsyn, S. K. Analysis of laser radiation using the Nonlinear Fourier transform. *Nature Communications* **10**, 5663 (2019).

39  Zhou, Y., Zhou, G., Qin, Y., Fu, S., Lau, A. P. T. & Wong, K. K. Y. Unveiling Laser Radiation of Multiple Optical Solitons by Nonlinear Fourier Transform. *Laser & Photonics Reviews* **17**, 2200731 (2023).

40  Li, H., Li, X., Zhang, S., Liu, J., Yan, D., Wang, C. & Li, J. Vector Staircase Noise-Like Pulses in an Er/Yb-Codoped Fiber Laser. *Journal of Lightwave Technology* **40**, 4391-4396 (2022).

41  Chang, S., Zheng, Y., Wang, Z. & Shen, C. Raman-scattering-assistant large energy dissipative soliton and multicolor coherent noise-like pulse complex in an Yb-doped fiber laser. *Opt. Lett.* **46**, 5695-5698 (2021).

42  Li, X., Zhang, S., Liu, J., Yan, D., Wang, C. & Yang, Z. Symbiotic coexistence of noise-like pulses. *Optics Express* **29**, 30449-30460 (2021).

43  Du, Y., He, Z., Zhang, H., Gao, Q., Zeng, C., Mao, D. & Zhao, J. Origin of spectral rogue waves in incoherent optical wave packets. *Physical Review A* **106**, 053509 (2022).

44  Yang, Z., Yu, Q., Wu, J., Deng, H., Zhang, Y., Wang, W., Xian, T., Huang, L., Zhang, J., Yuan, S., Leng, J., Zhan, L., Jiang, Z., Wang, J., Zhang, K. & Zhou, P. Ultrafast laser state active controlling based on anisotropic quasi-1D material. *Light: Science & Applications* **13**, 81 (2024).

45  Wei, C., Zhu, X., Norwood, R. A. & Peyghambarian, N. Passively continuous-wave mode-locked Er3+-doped ZBLAN fiber laser at 2.8 μm. *Opt. Lett.* **37**, 3849-3851 (2012).

46  Nie, M., Li, B., Jia, K., Xie, Y., Yan, J., Zhu, S., Xie, Z. & Huang, S.-W. Dissipative soliton generation and real-time dynamics in microresonator-filtered fiber lasers. *Light: Science & Applications* **11**, 296 (2022).

47  Jeong, Y., Vazquez-Zuniga, L. A., Lee, S. & Kwon, Y. On the formation of noise-like pulses in fiber ring cavity configurations. *Optical Fiber Technology* **20**, 575-592 (2014).

48  Herink, G., Kurtz, F., Jalali, B., Solli, D. R. & Ropers, C. Real-time spectral interferometry probes the internal dynamics of femtosecond soliton molecules. *Science* **356**, 50-54 (2017).

49  Lau, K. Y., Cui, Y., Liu, X. & Qiu, J. Real-time investigation of ultrafast dynamics through time-stretched dispersive fourier transform in mode-locked fiber lasers. *Laser & Photonics Reviews* **17**, 2200763 (2023).

50  Du, Y., He, Z., Gao, Q., Zeng, C., Mao, D. & Zhao, J. Intermediate state between a solitary singlet and a molecule in lasers. *Physical Review A* **107**, 053512 (2023).